\documentclass{camera}
\usepackage{epsfig}
\newcommand{\apm}[2]{\ensuremath{^{+#2}_{-#1}}}
\newcommand{\Ra}{\ensuremath{\Rightarrow}}
\def\etal{{\it et~al.}}%
\def\GeV{\ifmmode {\mathrm{\ Ge\kern -0.1em V}}\else
                   \textrm{Ge\kern -0.1em V}\fi}%
\def\MZ{\ensuremath{m_{\mathrm{Z}}}}%
\def\MW{\ensuremath{m_{\mathrm{W}}}}%
\def\Mt{\ensuremath{m_{\mathrm{t}}}}%
\def\MH{\ensuremath{m_{\mathrm{H}}}}%
\def\swsq{\ensuremath{\sin^2\!\theta_{\mathrm{W}}}}%

\def\qqbar{\ensuremath{\mathrm{q}\bar{\mathrm{q}}}}
\def\GW{\ensuremath{\Gamma_{\mathrm{W}}}}%
\def\ROb{\ensuremath{R_{\mathrm{b}}^{\mathrm{0}}}}%
\def\ROc{\ensuremath{R_{\mathrm{c}}^{\mathrm{0}}}}%
\def\AFBOl{\ensuremath{A_{\mathrm{FB}}^{0,\ell}}}%
\def\AFBOb{\ensuremath{A_{\mathrm{FB}}^{0,\mathrm{b}}}}%
\def\AFBOc{\ensuremath{A_{\mathrm{FB}}^{0,\mathrm{c}}}}%
\def\Al{\ensuremath{\cal{A}_\ell}}%
\def\Ab{\ensuremath{\cal{A}_\mathrm{b}}}%
\def\Ac{\ensuremath{\cal{A}_\mathrm{c}}}%
\def\delalp{\ensuremath{\Delta \alpha_{\mathrm{had}}^{(5)}}}%
\newlength{\leftcolwidth}%
\newlength{\rightcolwidth}%
\newcommand{\doublecolumn}[3]%
{%
\setlength{\leftcolwidth}{#1}%
\setlength{\rightcolwidth}{\textwidth}%
\addtolength{\rightcolwidth}{-1.0\leftcolwidth}%
\addtolength{\rightcolwidth}{-1.5ex}%
\begin{minipage}{\leftcolwidth}%
{#2}
\end{minipage}%
\begin{minipage}{1.5ex}%
\mbox{\,}
\end{minipage}%
\begin{minipage}{\rightcolwidth}%
{#3}
\end{minipage}%
}
\begin{document}

%
\title{THE GLOBAL \\ELECTROWEAK FIT}

%
\author{S. Villa}

%
\organization{University of California, Riverside}

\maketitle

\abstract{We report the results of the global electroweak
fit, with emphasis on the most recent results which served as 
input of the fit. The output of the fit sets also
limits on the Standard Model Higgs mass.}

\vspace{1cm}

%
\section{Introduction}\label{sec:intro}
%
The Standard Model of particle physics (SM) represents certainly
the biggest
success of 20th century physics. Its validity over a very wide
range of energies has been experimentally tested to unprecedented
precision, showing a perfect agreement of theory and experiment.
The use of electroweak corrections, combined with precision
measurements is the main strategy used to evaluate 
the parameters of the model which are
still unmeasured or which are measured with the poorest accuracy, 
such as the mass of the top, \Mt,  and of the Higgs, \MH.
Electroweak radiative corrections to physical observables 
are computable in perturbation theory and they depend 
quadratically on \Mt \ and logarithmically on \MH~\cite{yellowbook}.  

The global electroweak fit combines all the information coming from many
experiments into one single $\chi^2$ fit, to obtain the best
evaluation of all the parameters of the SM.
The fit accepts as input the following measurements from LEP:
the mass and width of the Z, the hadronic pole cross section of Z exchange, 
the Z leptonic branching ratio, the leptonic forward backward asymmetry 
(\AFBOl), the $\tau$ polarisation, the \qqbar \ charge asymmetry, and 
the mass and width of the W boson (\MW, \GW).
Other inputs come from the combination of SLD and LEP heavy flavour 
measurements: the ratios of b and c partial widths of the Z to its total
hadronic width (\ROb, \ROc), the b and c forward backward asymmetries and
the coupling parameters \Ab \ and \Ac. 
The other measurements used are
the coupling parameter \Al \ from SLD, \MW, \GW \ and \Mt \ from 
$\mathrm{p}\bar{\mathrm{p}}$ colliders, the measurements of Atomic
Parity Violation (APV), \swsq \ from $\nu$N scattering and the contribution
of light quarks to the photon vacuum polarisation (\delalp) from 
low energy $\mathrm{e}^+ \mathrm{e}^- \rightarrow \qqbar$.

In the following we will shortly review the inputs and the results
of the global electroweak fit performed in Winter 2002~\cite{EWWGwin02}.
%
\section{New and updated experimental inputs}
%
The most significant changes in the Winter 2002 global electroweak fit 
are the new results on \AFBOb \ and \AFBOc \ from Aleph, 
the final results from NuTeV, the inclusion of \GW \ in the fit, 
and the new interpretation of the APV experiments.
The  Aleph measurement of \AFBOb \ and \AFBOc \ using leptons is 
described in~\cite{AlephAFB}. The change induced by these results
on the electroweak averages are of +1/4 of standard deviation for
\AFBOb \ and of +2/3 of standard deviation for \AFBOc. 
The NuTeV experiment has presented a final analysis of their data~\cite{nutev}
of the scattering of $\nu_\mu$ and $\bar{\nu}_\mu$ on nuclei of iron.
\begin{figure}[!htb]
\begin{center}
\epsfig{file=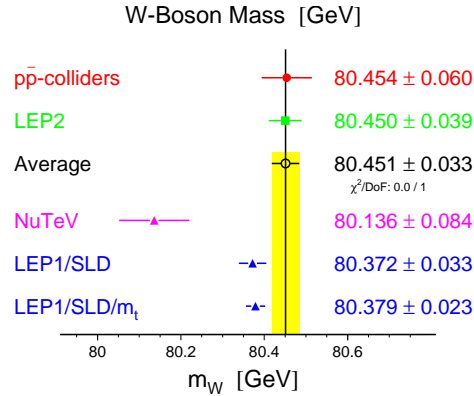, width=0.5\textwidth}
\parbox{0.85\textwidth}{
\caption{\label{fig:MW} Comparison of the W mass as measured by 
$\mathrm{p}\bar{\mathrm{p}}$-colliders and LEP2 experiments and
as derived by the NuTeV measurement and by LEP1 and SLD electroweak data.}
}
\end{center}
\end{figure}
The value of \swsq \ they obtain from the ratio of neutral to charged current
cross sections is:
\begin{eqnarray*}
\swsq & = & 0.2277 \pm 0.0013 (\mathrm{stat}) \pm 0.0009 (\mathrm{syst}) \: ,
\end{eqnarray*}
more than 3 standard deviations higher than the value
obtained by the combination of all the other available electroweak data.
In fig.~\ref{fig:MW} the value of \MW \ deduced by this measurement is
compared to the direct measurements of LEP2 and $\mathrm{p}\bar{\mathrm{p}}$ 
colliders and to the values derived from LEP1 and SLD electroweak 
measurements.
This discrepancy has so far not been given any satisfactory explanation, 
and it is therefore accepted as a statistical fluctuation.
For the first time the value of \GW \ measured by LEP and Tevatron
has been included in the electroweak global fit.
The combined measurement~\cite{GW} is $\GW = 2.13 \pm 0.07 \GeV$.
Finally, a new update of the measurement of the nuclear weak charge of
cesium~\cite{APV} has been included in the fit. The updated value is
$ Q_{\mathrm{W}}(\mathrm{Cs}) = -72.39 \pm 0.29 (\mathrm{exp}) \pm 0.51 (\mathrm{theo}) $, 
in good agreement with the SM expectation: 
$ Q_{\mathrm{W}}(\mathrm{Cs})^{SM} = -72.885 $.
%
\section{Results and conclusions}
%
The electroweak global fit is based on the SM predictions as implemented
in the ZFITTER~\cite{ZFITTER} and TOPAZ0~\cite{TOPAZ0} programs. It
accepts as input all the parameters that we have listed in sec.~\ref{sec:intro}
and gives as output estimates for all the parameters of the model, 
including the unmeasured ones such as \MH, the strong coupling constant 
$\alpha_s (\MZ^2)$ and the ones with the largest experimental
uncertainty, such as \Mt. 

\begin{figure}[!htb]
\doublecolumn{0.55\textwidth}{
\begin{center}
\vspace{0.6cm}
\epsfig{file=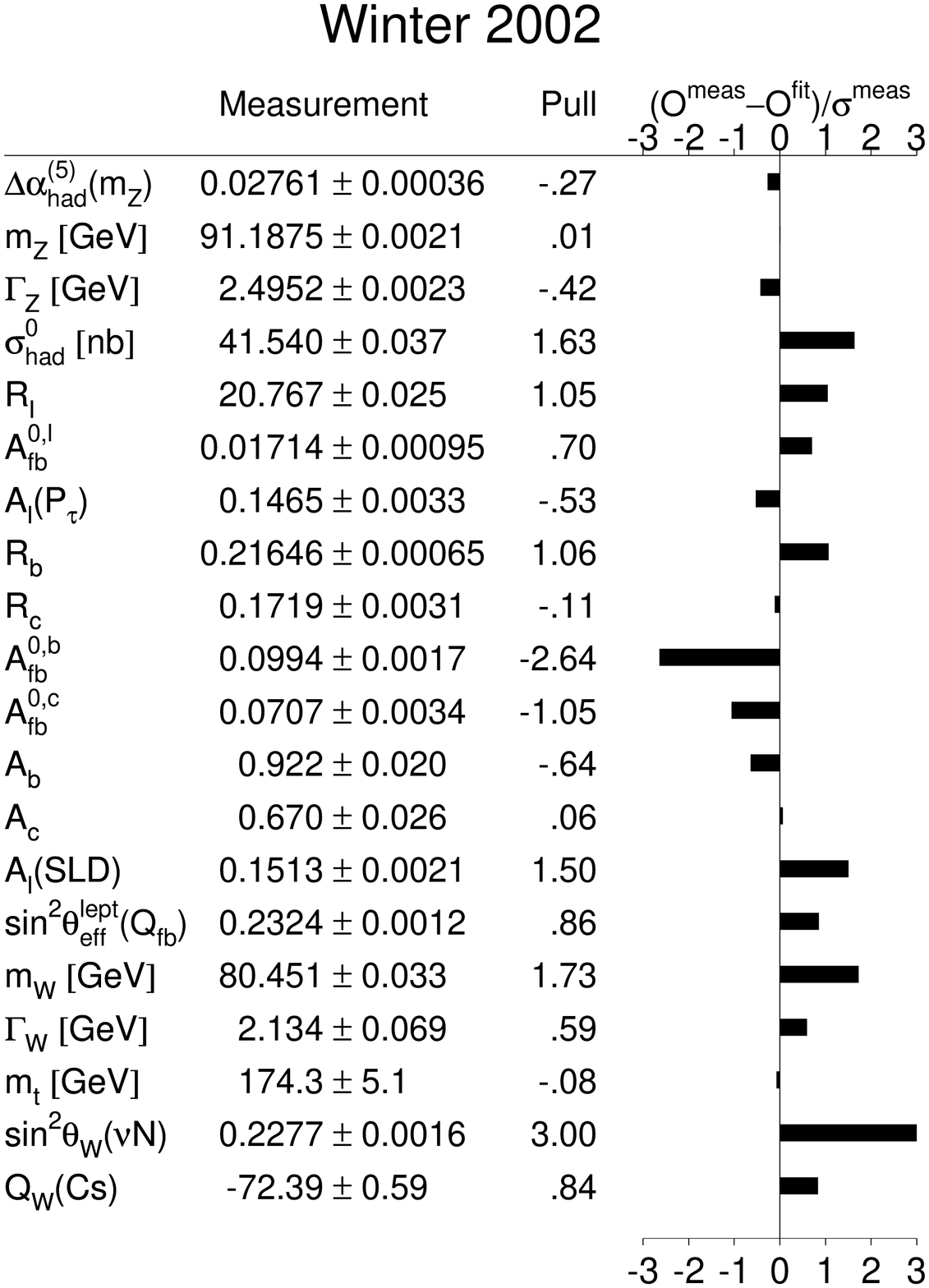, width=0.95\textwidth}
\parbox{0.9\textwidth}{
\caption{\label{fig:fit} Results of the electroweak global fit. Input
parameters are listed with their experimental value, and with 
the pull of the fit, defined by the difference between the measured and
fitted value divided by the experimental uncertainty.}
}
\end{center}
}
{
\vspace{3.4cm}
\begin{center}
\epsfig{file=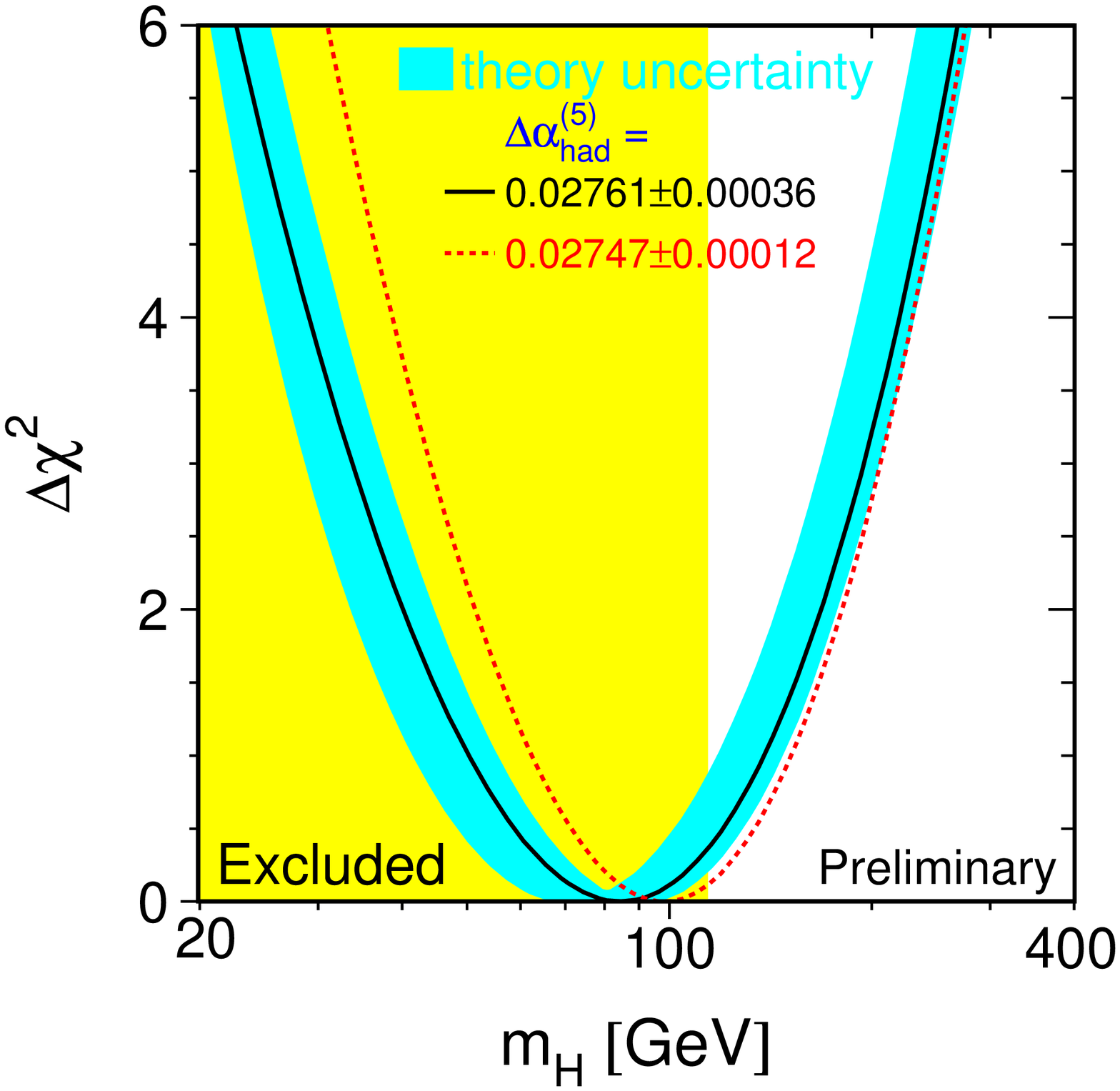, width=0.9\textwidth}
\parbox{0.8\textwidth}{
\caption{\label{fig:higgs} The $\Delta \chi^2$ curve for the fit
of the Higgs mass. The dark shaded band corresponds to the theoretical uncertainty, 
while the dotted line corresponds to a different evaluation of
\delalp~\cite{traconiz}. The mass range experimentally excluded by LEP searches
is represented by the light shaded area.}
}
\end{center}
}
\end{figure}

A summary of the fit results is shown in fig.~\ref{fig:fit}.
The largest pulls are given by the NuTeV result and
by \AFBOb. The $\chi^2 / \mathrm{d.o.f}$ of this fit is $28.8/15$,
corresponding to a probability of 1.7\%.
The fit repeated excluding the NuTeV results yields a 
$\chi^2 / \mathrm{d.o.f}$ of $19.9/14$ with probability 14.3\%.
%
%
The most interesting output of the fit is the estimate of the Higgs
mass. Figure~\ref{fig:higgs} shows the $\Delta \chi^2$ curve for
\MH; the shaded band correspond to the theoretical uncertainty.
The 1$\sigma$ estimate for the mass of the SM Higgs is
$ \MH  =  85 \apm{34}{54} \GeV$, with an upper limit at  95\%~CL
of 196 \GeV. This result is changed only by a few \GeV \ when the NuTeV result
is not used in the fit. 

To summarise, the SM describes very well all the data which is used
to perform the electroweak global fit. The two measured parameters which show
the largest disagreement with their expected values are the measurement of
\swsq \ from NuTeV and the b forward-backward asymmetry. It is 
possible to explain such disagreements as statistical fluctuations even though 
we cannot exclude that they represent hints of yet unknown physical processes
not described by the Standard Model.
%
%

%
%
\end{document}